\documentstyle[aps,prl,twocolumn]{revtex}
\input{epsf}
\begin{document}
\twocolumn[\hsize\textwidth\columnwidth\hsize\csname@twocolumnfalse\endcsname

\title {A secure key-exchange protocol with an absence of injective functions} 
\author {R. Mislovaty$^1$,
Y. Perchenok$^1$, I. Kanter$^1$ and W. Kinzel$^2$} \address{ $^1$
Minerva Center and Department of Physics, Bar-Ilan University,
Ramat-Gan 52900, Israel \\ $^2$ Institut f\"ur Theoretische Physik,
Universit\"at W\"urzburg D-97074, Germany}

\maketitle
    
\begin{abstract}    
The security of neural cryptography is investigated.  A key-exchange
protocol over a public channel is studied where the parties exchanging 
secret messages use multilayer neural networks which are trained by 
their mutual output bits and synchronize to a time dependent secret 
key. The weights of the networks have integer
values between $\pm L$. Recently an algorithm for an eavesdropper
which could break the key was introduced by Shamir et al.\cite{adi}.
We show that the synchronization time increases with $L^2$ while the
probability to find a successful attacker decreases exponentially with
$L$. Hence for large $L$ we find a secure key-exchange protocol which
depends neither on number theory nor on injective trapdoor functions
used in conventional cryptography.
\end{abstract} 
]   

The ability to build a secure channel is one of the most challenging
fields of research in modern communication\cite{1}.   
One of the fundamental tasks of
cryptography is to generate a {\it key-exchange protocol}. Both
partners start with private keys and transmit -- using a public
protocol -- their encrypted private keys which, after some
transformations, leads to a common secret key.  A prototypical
protocol for the generation of a common secret key is the
Diffie-Hellman key exchange protocol\cite{1}.

All known secure key-exchange protocols use one-way functions, 
which are usually 
based on
number theory and in particular on the difficulty in factorizing a
product of long prime numbers\cite{1,2}.  Typically, $N$ bits -- the
length of the key -- are transmitted between the two partners and
transformed by an injective function to the common key. This function
usually can be inverted by a secret trapdoor.  One of the fundamental
questions in the theory of cryptography is firstly whether it is
possible to build a secure cryptosystem which does not rely on number
theory, secondly, whether one can transmit less than $N$ bits and
thirdly, whether one can generate very long keys which can be directly
used for one-time stream ciphers\cite{1}.

In our recent paper\cite{epl} we presented a novel principle of a
key-exchange protocol based on a new phenomenon which we observed for
artificial neural networks. The protocol is based on the
synchronization of feedforward neural networks by mutual learning. It
was shown by simulations and by the analytical solution of the
dynamics that synchronization is faster than the learning of a naive
attacker that is trying to reveal the weights of one of the
parties\cite{epl,michal}. Our new approach does not rely on previous
agreement on public information , and the only secret of each one of
the parties is the initial conditions of the weights. The protocol
generates permanently new keys and can be generalized to include the
scenario of a key-exchange protocol among {\it more than two
partners}\cite{epl}. Hence, we suggest a symmetric
key-exchange protocol over a public channel which simplifies the task
of key management. 
The parties 
exchange a finite number of bits less than $N$ and can generate 
very long keys by fast calculations.\cite{comment1}

This protocol for the given parameters in \cite{epl} ($K=L=3$) was
recently shown to be breakable by an ensemble  of advanced
flipping attackers\cite{adi}. In such an ensemble, there is a 
probability that a low percentage of the attackers will find the
key. Someone reading all the decrypted messages will determine
the original plaintext from the message which has a meaning.
This result raises the question of the existence of a
secure key-exchange protocol based on the synchronization of neural
networks.

In this Letter we demonstrate that the security of our key-exchange
protocol against the flipping attack increases as the synchronization
time increases.  The mechanism used to vary the synchronization time is
the depth of the weights, i.e. the number of values for each
component of the synaptic weights.  The main result in this Letter
is that with increasing depth the probability of an attacker finding
the key decreases exponentially with the depth.  Hence we conjecture
that a key-exchange protocol exists in the limit where the
synchronization time diverges.  We also present a variant of our
original scheme which includes a permutation of a fraction of the
weights. 

In our original scheme each party of the secure channel, $A$ and $B$,
is represented by a two-layered perceptron, exemplified here by a
parity machine (PM) with $K$ hidden units. More precisely, the size of
the input is $KN$ and its components are denoted by
$x_{kj},~k=1,~2,~...,~K$ and $j=1,~...,~N$. For simplicity, each input
unit takes binary values, $x_{kj} =\pm1$.  The $K$ binary hidden units
are denoted by $y_1,~y_2,~...,~y_K$. Our architecture is characterized
by non-overlapping receptive fields (a tree), where the weight from
the j$th$ input unit to the k$th$ hidden unit is denoted by $w_{kj}$,
and the output bit $O$ is the product of the state of the hidden
units.  The weights can take integer values bounded by $|L|$, i.e., $
w_{kj}$ can take the values $-L,~-L+1,~...,~L$.

The secret information of each of the parties is the initial
value for the $2KN$ weights, $w_{kj}^A$ and $w_{kj}^B$.  The parties
do not know the initial weights of the other party which are used to
construct the common secret key.
     
Each network is then trained with the output of its partner.  At each
training step a new common public input vector $(x_{kj})$ is needed
for both parties.  For a given input, the output is calculated in the
following two steps.  In the first step, the state of the $K$ hidden
units, $y^{A/B}_k$ of the two parties, are determined from the
corresponding fields
\begin{equation}     
y^{A/B}_k = \mbox{sign} \lbrack      
\sum\limits^N_{j=1} \; w^{A/B}_{kj} \; x_{kj} \rbrack     
\end{equation}     
    
\noindent In the case of zero field,  $\sum w^{SA/B}_{kj} \; x_{kj}
=0$, $A/B$ sets  the hidden unit to $1/-1$.   In the next step the
output $O^{A/B}$ is determined by the  product of the hidden units,
 $O^{A/B}= \Pi_{m=1}^K y^{A/B}_m$.  The output bit of each party is
transmitted to its partner. In the event of disagreement, $O^A \ne O^B$,
the weights of the  parties are updated according to the following
Hebbian  learning rule\cite{5,5a}
\begin{eqnarray}     
\mbox{if} \ \Big( O^{A/B} y^{A/B}_k > 0 \Big) \; & \mbox{then} & \;
w^{A/B}_{kj} = w^{A/B}_{kj} +  O^{A/B} \, x_{kj} \nonumber \\ \mbox{if}
\ \Big( |w_{kj}^{A/B}| > L \Big)\; \; & \mbox{then} & \; w_{kj}^{A/B}
= \mbox{sign}(w_{kj}^{A/B}) \ L
\label{two}    
\end{eqnarray}     
Only weights belonging to hidden units which are in the same state as
their output unit are updated.  Note that from the knowledge
of the output, the internal representation of the hidden units is not
uniquely determined because there is a $2^{K-1}$ fold degeneracy. As a
consequence, an attacker cannot know which weight vectors are
updated according to equation (2). 
Nevertheless, although parties A and B do not have 
more information than an attacker, they still can synchronize.

The synchronization time is finite even in the thermodynamic
limit\cite{epl,michal}.  For $K=L=3$, for instance, the
synchronization time $t_{av}$ converges to $\simeq 400$ for
large networks.  This observation was recently confirmed by an
analytical solution of the presented model\cite{michal}. Surprisingly,
in the limit of large $N$ one needs to exchange only a few hundred 
bits to obtain agreement between $3N$ components.\cite{comment1,signature}

An attacker eavesdropping on the channel knows the algorithm as well
as the actual mutual outputs, hence he knows in which time steps the
weights are changed.  In addition, an attacker knows the input
$x_{kj}$ as well.  However, the attacker does not know the initial
conditions of the weights of the parties and as a consequence, even
for the synchronized state, the internal representations of the hidden
units of the parties are hidden from the attacker. As a result he does
not know which are the weights participating in the learning step.
Note that for random inputs all $2^{k-1}$ internal representations
appear with equal probability at any stage of the dynamical process.
The strategy of a naive attacker which has the same architecture as
the parties is defined as follows\cite{epl}. The attacker tries to
imitate the moves of one of the parties, $A$ for instance.  The
attacker is trained using  its internal representation, the input vector
and the output bit of $A$, and the training step is performed only if
$A$ moves (disagreement between the parties). Note that the trained
weights of a naive attacker are only weights belonging to hidden
units that are in agreement with $O^A$. Simulations as well as
analytical solution of the dynamics indicate that the learning time of
a naive attacker is much longer than the synchronization
time\cite{epl,michal}. Hence our key-exchange protocol is robust
against a large ensemble of naive attackers.

Recently, an efficient flipping attack was presented\cite{adi}. The
strategy of a flipping attacker, $C$ is as follows.  In the event of a
disagreement between the parties, $O^A \ne O^B$ and $O^C =O^A$, the
attacker moves as for the naive attack following its internal
presentation, the common input and $O^A$.  In the case where the parties
move but the attacker does not agree with $A$, $O^A \ne O^B$ and $O^C
\ne O^A$, the move consists of the following two steps. In the first
step the attacker flips the sign of one of its $K$ hidden units {\it
without altering the weights}.  The selected hidden unit is 
$K_0$ with the minimal absolute local field
\begin{equation}
K_0 =min_m(|h_m^C|)
\end{equation}
where $h_m^C$ is the local field on the $m$th hidden units of the
attacker (see eq. (1) for the definition of the local field). After
flipping one hidden unit the new output of the attacker
agrees with that of $A$. The learning step is then performed with the
new internal presentation and with the strategy of the naive attacker.
The flipping attack is based on the strategy that a flipping attacker
develops some similarity with the parties. This similarity can be
measured by the fraction of equal weights which is greater than
$1/(2L+1)$, a result for a random attacker, or by a positive overlap
between the weights of $C$ and $A$\cite{michal}. The minimal change in
the weights which preserves the already produced similarity with $A$
and which is also consistent with the current input/output relation is
most probable by changing the weights of the hidden units with the
minimal absolute local field.  Simulations as well as the analytical
solution of the dynamics of the flipping attackers\cite{michal1}
indicate that there is a high probability that there is a successful attacker
among a few dozen attackers. By a successful attacker we mean
an attacker with a learning time smaller than the synchronization time
between the parties.  This attacker achieves the same weights as for
$A$ before the synchronization process terminates.  In Fig. 1 the
average synchronization time, $t_{av}$, as well as its standard
deviation as a function of $L$ for $K=3$ and $N=10^3$ are
presented. Results were averaged over $\sim 10^4$ different runs,
where each run is characterized by different initial conditions for
the parties and a different set of inputs. Results indicate that the
synchronization time increases as $L^2$, for $L < O(\sqrt{N})$.  This
scaling is consistent with the analytical solution of
ref\cite{metzler} where for $L=\sqrt{N}, ~t_{av} \propto N$.  
For $L = O(\sqrt{N})$ we observe in simulations a crossover to the
scaling behavior $t_{av} \propto \sqrt{N}L$. This crossover explains
the deviation of $t_{av} \propto L^{\sigma}, \sigma=1.91 <2$ (see
Fig. 3), and furthermore $\sigma$ is expected to increase with $N$
(see Fig. 4).

\vspace{0.60cm}
\begin{figure}    
\centerline{\epsfxsize=3.25in \epsffile{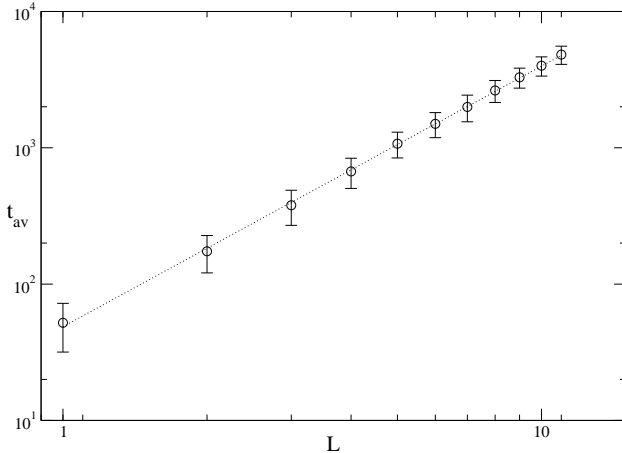}}    
\caption{The average synchronization time, $t_{av}$, and its
standard deviations as a function of $L$ for $K=3$ and $N=10^3$.  The
regression fit for the dotted line is $\sim 50L^{1.91}$.  
}
\end{figure}

\vspace{0.60cm}
\begin{figure}     
\centerline{\epsfxsize=3.25in \epsffile{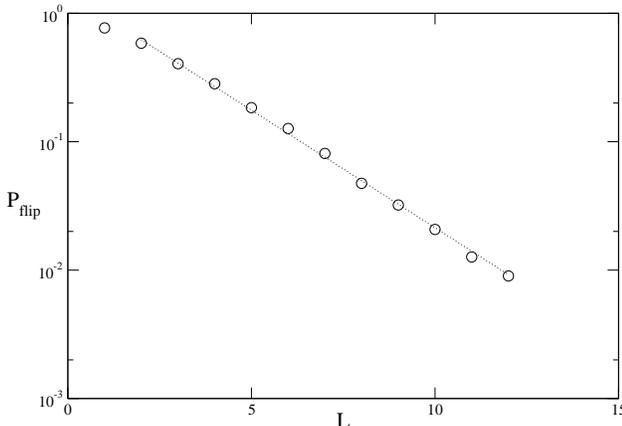}}    
\caption{The fraction of successful flipping attackers, $P_{flip}$, as a
function of $L$ for $K=3,~N=10^3$. The regression fit for the dotted
line is $\sim 1.4e^{-0.41*L}$ }
\end{figure}

In Figure 2 the fraction of
successful flipping attackers, $P_{flip}$, is presented as a function of
$L$. In order to reduce fluctuations in our simulations we define a
successful attacker as one which has $0.98$ fraction of
correct values for the weights at the synchronization time between the
parties.  Fig. 2 indicates that the success rate drops exponentially
with $L$.  To conclude, for $1 \ll L \ll \sqrt{N}$ the synchronization
time diverges polynomially while the probability of a successful
attacker drops exponentially. Hence for large $L$ our construction is
robust against the flipping attack 
(Practically, for $L \sim 85$ and $N > 2\cdot10^4$, the complexity 
of an effective flipping attack is greater than $2^{80}$).

Finally we note that the complexity of the
synchronization process for $1 \ll L \ll \sqrt{N}$ is $O(L^2N\log(N))$.
The factor $\log(N)$ is a result of a typical scenario of an
exponential decay of the overlap in the case of discrete
weights\cite{michal}. Hence, the complexity for the generation of a
large common key, $N \rightarrow \infty$, scales as $O(\log N)$
operations per weight. 
\vspace{0.60cm}
\begin{figure}     
\centerline{\epsfxsize=3.25in \epsffile{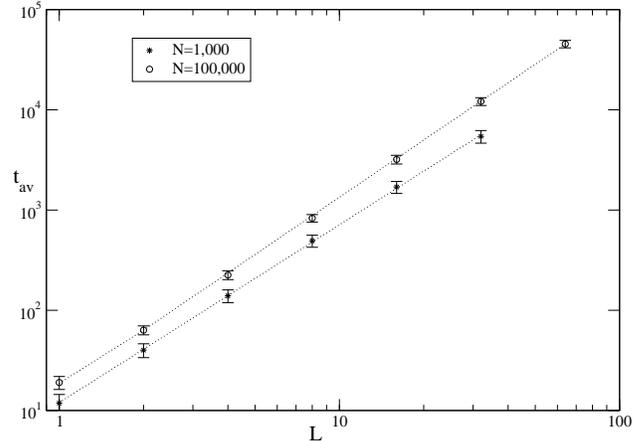}}
\caption{The learning time for a perceptron as a function of $L$ and
$N=10^3,~10^5$.  The regression power-law fit for $N=10^3,~10^5$ is
$\sim 12L^{1.77},~\sim 17L^{1.9}$, respectively.  }
\end{figure}

Let us compare now the complexity of an exhaustive attack with the
complexity of the flipping attack. For each input/output pair there
are $4$ possible configurations of the hidden units.  Hence to cover
all possible training processes over a period $t$ one has to deal with
an ensemble of $4^t$ scenarios. The crucial question is the scaling of
the minimal necessary period $t_0$ with $L$ which ensures a
convergence with the weights of party $A$. Since one of the attackers
among $4^{t_0}$ has an identical series of internal representations to
party $A$, the problem is reduced to calculating the weight vector
of a single perceptron. The learning time as a function of $L$ for a
perceptron attacker $K=1$ is presented in Fig. 3, indicating that for
large $N$, $t_0 \sim L^2$, as expected from similar analytically
solvable models\cite{metzler}. Hence the complexity of an exhaustive
attack scales exponentially with $L^{2}$ while for the flipping
attack the complexity is reduced to scale exponentially only with $L$.

In the following we show that one can increase the security of our
key-exchange protocol by the following variant of our dynamical
rules. The new ingredient is a permutation of a fraction $f$ of the
weights, and the protocol is defined by the following steps. In the case
where the parties move, we assign for each hidden unit a permutation
consisting of $F=fN$ pairs.  Each pair consists of a random selection of
two indices among $N$ of the trained hidden unit\cite{global}. The
three permutations for the three hidden units (which differ from step to
step) are part of the public protocol. In the case where a hidden unit is
trained we apply the assigned permutation for this hidden unit.  Note
that the permutations is an ingredient that prevents an attack where one
may assign for each weight (among $3N$) a probability equal to
one of the $2L+1$ possible values. During the dynamics one may try to
sharpen this probability around one of the possible
values\cite{adi}. The permutations are responsible for mixing these
probabilities as a function of time.

Results indicate that there are two different scaling behaviors for
$t_{av}(L)$ and $P_{flip}(L)$ as a function of the total number of
permuted pairs, $M$, during the synchronization process.  As long as
$M < \phi KN$ where $\phi \sim 1 $, the permutations do not affect the
synchronization time, $t_{av}(L)= AL^2$; $A \sim 60$ is independent of
the permutations ($A$ increases slightly with $N$ and is asymptotically
expected to scale with $\log(N)$\cite{michal}). This scaling behavior
can be observed for $L < \sqrt{3\phi N/(60f)}$. Hence in order to
observe the scaling, $t_{av} \sim 60L^2$ over a decade of $L$ one has
to choose a large $N$ and a very small $F$. In Fig. 4 the average
synchronization time, $t_{av}$, and its standard deviations as a
function of $L$ are presented for $K=3$, $N=10^5$ and $F=0,~3$ (number
of permuted pairs is $3$ per hidden unit). An insignificant deviation
from the scaling behavior is observed only for $L \ge 32$.  In the
inset of Fig. 4, similar results are presented for $N=10^3$ with
$F=3$, and $N=10^4$ with $F=3$ and $20$. The deviation from the
scaling behavior is observed for a larger $L$ as we increase $N$ or as
we decrease $F$ ($L < \sqrt{3\phi N/(60f)}$).  We also measured
$P_{flip}(L)$ $L<10$ for $N=10^4,~10^5$ with $F=3$ or $F=0$. We
realized that $P_{flip}$ is independent of $F$ and it decreases
exponentially with $L$.  The permutations do not affect the
exponential drop, $P_{flip} \propto e^{-BL}$, where $B$ appears to
increase with $N$. Note that although the permutations do not affect
$t_{av}$ and $P_{flip}$, the accumulated affect of the permutations
over all the synchronization process is significant.  In the event
that the flipping attacker does not use the permutation, a dramatic
drops in $P_{flip}$ is observed\cite{michal1}. The analysis of the
scaling behavior of $t_{av}$ and $P_{flip}$ in the second regime $L >
\sqrt{3\phi N/(60f)}$ is beyond our computational ability, where huge
fluctuations are observed.

The scaling of $P_{flip}$ may be examined against other classes of
attacks including a genetic attack, a majority attack and a
flipping attack where the weights of the selected hidden unit are
modified to actually flip the sign of the hidden unit\cite{adi}. Our
results indicate that all such types of attacks are less
efficient than the flipping attack presented.  
Hence, for all known attacks  neural cryptography is 
secure in the limit of large values of $L$.

We thank Adi Shamir for critical comments on the manuscript.

\vspace{0.60cm}
\begin{figure}     
\centering{\epsfxsize=3.25in \epsffile{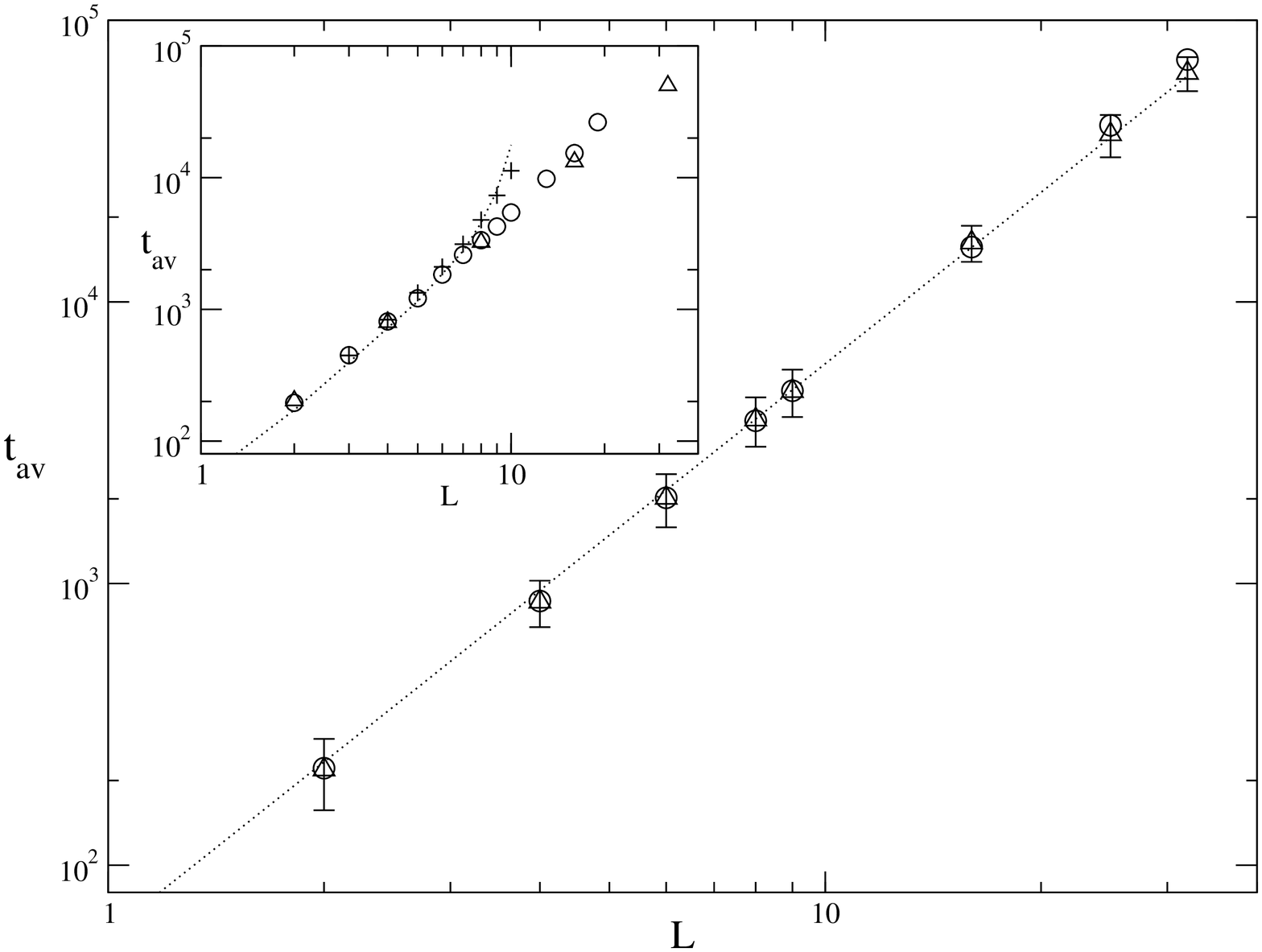}}
\caption{The synchronization times, $t_{av}$, and their standard
deviations as a function of $L$ for $K=3$, $N=10^5$ with $F=0$
($\bigtriangleup$) and $F=3$ ($\bigcirc$). The regression fit for $2
\le  L \le 25$ , dotted line, is $\sim 57.3L^{2.02}$.  Inset: $t_{av}$
as a function of $L$, $N=10^3$, $F=3$ (dashed line), $N=10^4$
$F=0,~3,~20$ ($\bigtriangleup, \bigcirc, +$).  }
\end{figure}


\begin{thebibliography}{10}
\bibitem{comment1}However, one should keep in mind that the two partners
do not learn the initial weights of each other, they are just
attracted to a dynamical state with opposite weights.
\bibitem{5a} A stationary synchronization
state of antiparallel weights for the
parties can be modified to the dynamical rules where weights are
updated only in the case of an agreement between the parties. The
stationary solution  in this case is parallel weights for the parties.
\bibitem{signature} Note that from the synchronized weights it is
difficult to determine initial set of weights consistent with the
sequence of transmitted bits. This property may serve for other tasks
of the secure channel such a one-time pad signature.
\bibitem{global} Similar results were also obtained for global permutations,
where the second index of each pair was randomly selected from the other two
hidden units.
\end{thebibliography}
\end{document}